\documentclass[a4paper,twoside,10pt]{article}
\usepackage{graphicx,publaob}

\def\papertype{Invited paper}
\begin{document}

\title{COSMIC RAY ORIGIN: BEYOND THE STANDARD MODEL(S). THE CASE OF PULSAR WIND NEBULAE AND UNIDENTIFIED VERY HIGH ENERGY GAMMA-RAY SOURCES.}

\authors{O. TIBOLLA}

\address{$^1$Mesoamerican Centre for Theoretical Physics (MCTP), Universidad Aut\'onoma de Chiapas (UNACH), Carretera Emiliano Zapata Km. 4, Real del Bosque (Ter\'an). 29050 Tuxtla Gutierrez, Chiapas, Mexico.}
\Email{omar.tibolla}{gmail.com}{}

\markboth{COSMIC RAY ORIGIN: BEYOND THE STANDARD MODEL(S)}{O. TIBOLLA}

\abstract{The riddle of the origin of Cosmic Rays is open since one century.
Recently we got the experimental proof of hadronic acceleration in Supernovae Remnants, however new questions rised and no final answer has been provided so far.
Gamma ray observations above 100 MeV reveal the sites of cosmic ray acceleration to energies where they are unaffected by solar modulation.
In the last years the knowledge in this field of research widely increased, however almost 50\% of the TeV ($> 10^{12}$ eV) Galactic sources are still unidentified; at GeV ($> 10^9$ eV) energies, 67\% of EGRET sources were unidentified and also with the newer generation of gamma-ray satellites we have the same result: in fact, at low Galactic latitudes (b$<$10 deg), 62\% of the Fermi LAT detected sources have no formal counterpart.
Hence understanding the high energy unidentified sources will be a crucial brick in solving the whole riddle of Cosmic Rays origin.
Several examples will be shown, underlining the importance of the so-called ''dark sources``.
Both theoretical aspects (with particular emphasis to the so-called Ancient Pulsar Wind Nebulae scenario) and their observational proofs will be discussed.
}

\section{Cosmic Ray Origin: Standard Model(s) and beyond}

In this short article we are summarizing one of the oldest questions in astrophysics: where the cosmic rays (CRs) come from? In which astrophysical objects are they accelerated?
Entire international conferences have been devoted to try to answer this question, such as the two editions of the ``Cosmic Ray Origin - beyond the standard models'' international conference (CRBTSM \footnote{ http://crbtsm.eu }); this answer is long and can be complex (e.g. Tibolla \& Drury 2014).  

CRs have been discovered more than one century ago with different experimental procedures.
Their discovery in water (in the Bracciano lake and in the Tirreno sea) in the years 1907-1912 can be attributed to the Italian physicist Domenico Pacini (Pacini 1912).
Their discovery utilizing an electroscope on the Eiffel Tower can be attributed to the German gesuit physicist Theodor Wulf (Wulf 1909), who was also the first who spoke about ``Hoehenstrahlung'' (i.e. ``radiation from above'').
The invention of electroscopes by Wulf led to much more precise experiments, i.e. the balloon flights at 5 km of altitude by the Austrian physicist Victor Hess (Hess 1912) and at 9 km of altitude by the German physicist Werner Heinrich Gustav Kolh\"orster (Kolh\"orster 1913). 
In particular, the first flight by Hess convinced everybody that CRs are an astrophysical phenomenon (and in 1936 Hess was awarded by the Nobel prize for CRs discovery) or, better, almost everybody.
In fact, the U.S. physicist Robert Andrews Millikan was continuing being skeptical about the conclusions of his European colleagues and disputed them for several years, before confirming their discovery; ironically the term ``cosmic rays'' originated with Millikan.
All in all, more than one century after their discovery, we still ignore the origin of CRs.

Since the discovery of the Supernovae (SN) phenomenon, a connection with CRs has been suggested (Baade \& Zwicky 1934). 
The terms ``standard picture'' or ``standard model'' generally refer to the majestic mongraph book of Ginzburg and Syrovatskii (Ginzburg \& Syrovatskii 1964).
In order to summarize extremely much their conclusions, we could say that according to the standard model: primary Galactic CRs up to energies of the so called ``knee'' at about $10^{15}$ eV, are
accelerated in Supernovae Remnants (SNRs) shells. One expects about one supernova event every 30$-$50 years, and, in order to account for the energy density of CRs (about 1 eV/cm$^3$) and the CRs
confinement time deduced from spallation, the typical non-thermal energy release per supernova has to be about $10^{50}$ ergs, which is about 10\% of the total energy released in the SN explosions; in other words at least 10\% of the kinetic energy of the SN explosion has to go into CRs acceleration.

This idea was significantly strengthened at the end of the eighties by the fact that the prediction of the ``standard model'' seemed in astonishing good agreement with the typical amount of energy predicted to be
produced during the acceleration of relativistic particles in SNR shocks (V\"olk and Biermann 1988, Drury et al. 1989, Drury et al. 1994).

The detection by the Imaging Atmospheric Cherenkov Telescopes (IACTs) of TeV gamma-rays from SNRs spatially coincident with the sites of non-thermal X-ray emission, has strengthened the hypothesis of
the ``standard'' picture of CR origin for up to the ``knee'' energies; the most remarkable examples of it are given by the SNRs RX J1713.7-3946 (Aharonian et al. 2006) and RX J0852.0-4622 (Aharonian et al. 2005), the so-called Vela Jr SNR.

However the TeV gamma ray signal, can be explained in two ways:

\begin{itemize}
 
 \item{Inverse Compton scattering of relativistic electrons/positrons on background photons (CMB, infrared, X-rays, etc.); i.e. leponic models.} 
 
 \item{Neutral pion decay (due to proton-proton inelastic interactions); i.e. hadronic models.} 
 
\end{itemize}

Moreover hadronic and leptonic models are basically indistinguishable at TeV gamma-rays (e.g. Aharonian \& Atoyan 1999).
It is important to underline here that the hadronic models have to be the focus of our researches: in fact protons are more than 90\% of the total amount of CRs. SNRs proved to be sources of CR electrons decades ago, because of
the detected radio and X-ray emissions; what is missing here is the  compelling evidence for the acceleration of hadrons in SNRs.
However, once we fix the TeV gamma-ray energy spectra by means of IACTs observations, we can note that hadronic and leptonic models should have very different signatures in the adjacent GeV gamma-ray band (e.g. Aharonian \& Atoyan 1999).
So, utilizing a GeV gamma-ray observatory, we should be able to finally prove or disprove the ``standard picture'' of CRs origin.
And this study was indeed among the main scientific purposes (GLAST Science Brochure 1996) of \emph{Fermi-LAT}, that was built (also) to be a great observatory for SNRs (e.g. Tibolla 2007).
In fact Fermi-LAT detected several SNRs (and SNR/Molecular Clouds (MCs) interactions): e.g. Cygnus Loop (Katagiri et al. 2011), RX J0852.04622 (Tanaka et al. 2011), Tycho (Giordano et al. 2012), W51C (Abdo et al. 2009), W44 (Abdo et al. 2010), Cassiopeia A (Abdo et al. 2010b), RX J1713.7−3946 (Abdo et al. 2011), G8.70.1 (Ajello et al. 2012), W28 (Abdo et al. 2010c), W49B (Abdo et al. 2010d).

In particular, in the \emph{Fermi-LAT} SNRs sample, a very important discovery is represented by Tycho SNR.
In fact, in the case of Tycho, leptonic models are basically disproved; i.e. Tycho represents the ``smoking gun'', the ``hadronic fingerprint'', i.e. the answer to a 60-100 years old question about the origin of CRs (Giordano et al. 2012).
Moreover the efficiency of CRs acceleration (i.e. the percent of kinetic energy of the SN explosion which have to go into CRs acceleration) in Tycho SNR is more than the 10\% needed to confirm the ``standard'' model: e.g. Slane et al. 2014 calculated it to be $\sim$16\%.
So everything would look solved, the standard model confirmed and every question answered; but it is not so.

In addition, with \emph{Fermi-LAT} we could detect the characteristic neutral pion ``shoulder'' around 100 MeV for IC 443 and W44 (Ackermann et al. 2013).  
However, in my opinion, these two exceptional measurements, which are indeed very important (especially under a particle physics point of view), are less relevant than Tycho SNR since they deal
with SNRs/MCs interacting systems, in which hadronic accelerations are quite obvious and unavoidable, and not with ``naked SNRs'', which would instead go straight to the core of the CRs origin riddle.
Nevertheless these two discoveries seem to underline that everything look solved, the standard model confirmed and and every question answered.

Instead \emph{Fermi-LAT} observations on one of the most promising targets to confirm the standard model of CRs origin seems to go in a totally opposite direction. 
In fact, thanks to IACTs discoveries, before  \emph{Fermi-LAT} launch in 2008, the SNR RX J1713.7-3946 was considered one of the most promising targets and moreover one of the best threated by theoretical models (e.g. the majestic theory by Berezhko \& V\"olk 2006).
\emph{Fermi-LAT} observations showed that leptonic models are fitting very well the energy spectrum of RX J1713.7-3946, while hadronic models seem disproved.
This is considered one of the deepest problems for the standard model after \emph{Fermi-LAT} observations, however, in my point of view, there is a bigger issue, the so-called ``efficiency problem''.
In fact, also the SNRs that look more likely hadronic (and indeed in most of the \emph{LAT} detected SNRs, hadronic models are favored, even if leptonic ones cannot be fully discarded so far) do not seem efficient enough to reach the 10\% mentioned above.
On this regards, let us take Cassiopeia A (i.e. together with the Crab, one of the two most powerful explosions in our side of the Galaxy) as an example: even assuming that the whole GeV and TeV gamma-ray spectrum would be originated by hadronic processes, the total energy of the CRs accelerated in Cas A would correspond to $\sim$2\% of the kinetic energy of the initial SN explosion (e.g. Abdo et al. 2010b; Tibolla et al. 2011): far too few to support the standard model of CR origin.

Hence the one century old question about CRs orging is still open and we basically have two ways to proceed:

\begin{itemize}
 
 \item{we can modify the ``standard model'' (e.g. Vink 2012). However such modifications could even be drastic and lead to exotic solutions: e.g. what if the local density of CRs (i.e. the density inside the local bubble) is different
from the CRs density in the rest of the Galaxy, outside the local bubble? This would change totally the vision of the problem.} 
 
 \item{we can search alternative sources which might help us in closing the gap (if any), such as binary systems (e.g. Bednarek et al. 2014), protostellar jets (e.g. Araudo \& del Valle 2014), Galactic center (e.g. Thoudam 2014), Pulsar (e.g. Kotera 2014) and Pulsar Wind Nebulae (e.g. Weinstein 2014; Tibolla \& Drury 2014).
 However we must keep in mind that extragalactic sources, such as gamma-ray bursts (e.g. Meszaros 2014) and active galactic nuclei (e.g. Mannheim 2014), play a role as well.} 
 
\end{itemize}

and I personally think that we should actually do both.

A very natural support could come from Pulsar Wind Nebulae (PWNe).
In fact, looking the high energy sky (i.e. the ``non-thermal sky''), the dominant population is not represented by shell-type SNRs. Among the known gamma-ray sources, PWNe are the most
numerous. And indeed it was proposed that at the termination shock of the Pulsar Wind, also hadrons could be accelerated as well as leptons (e.g. Bednarek \& Protheroe 1997; Atoyan \& Aharonian 1996;
Cheng et al. 1990; Bednarek \& Bartosik 2003).
Moreover, since in general the total energy of PWNe is comparable with the one of SNRs (or, at least, we can easily reach $\sim$10\% of it), if hadrons can really be accelerated at the termination shock of Pulsar wind, this could indeed help the
global picture of CR origin.

But the most numerous (by far) population in absolute terms is represented by Unidentified Galactic sources. In fact almost 50\% of the TeV Galactic sources are
still unidentified. At GeV energies, this percent is even increasing; in fact 67\% of EGRET sources were unidentified (Hartman et al. 1999)
and also with the newer generation of gamma-ray satellites we reach the same result: in fact, at low Galactic latitudes (b $<$ 10 deg), 62\% of the Fermi
LAT detected sources have no formal counterpart (Ackermann et al. 2012).
Hence understanding the high energy unidentified sources could be a crucial brick in solving the whole riddle of CRs origin. Moreover the correlation between unidentified Galactic sources and
PWNe could be very close, at least for the so called ``dark sources''.




\section{Unidentified Very-High Energy gamma-ray sources}

With the famous H.E.S.S. Galactic Plane Survey (Aharonian et al.2006) of 2004/2005, many new TeV gamma-ray sources have been discovered; however $\sim$50\% of the newely discovered sources were unidentified: 
e.g. HESS J1614-518, HESS J1616-508, HESS J1632-478, HESS J1634-472, HESS J1702-420, HESS J1708-410, HESS J1745-303 and HESS J1837-069.

In the meaning while H.E.S.S. collaboration, in 2006-2017, extended the successful Galactic survey of 2004/2005, discovering a number of new gamma-ray emitting sources, many of which are unidentified;
VERITAS, another Imaging Atmospheric Cherenkov Telescope (IACT) observatory, in 2009-2017, started the scan of the northern part of the Galactic Plane.
Moreover the ``water Cherenkov technique'' is reaching, with the HAWC (High Altitude Water Cherenkov) experiment, sensitivities comparable to IACTs ones (however reaching
higher energies); Water Cherenkov experiments have the advantage to be ``full sky'' experiments and to not have limited duty cicles (i.e. observations are not stopped during the day or for not favorable weather conditions), so the
number of sources is going to increase (e.g. the 2HAWC catalog; Abeysekara et al. 2017).
Anyway the ratio does not seem to change: $\sim$50\% of the TeV Galactic gamma-ray sources seem still unidentified (e.g. Tam et al 2010).

The ``zoo'' of Galactic unidentified TeV gamma-ray sources is pretty various: 
\begin{itemize}

\item we have the so-called ``dark sources'', i.e. sources which do not show any kind of counterparts at lower energies (such as HESS J1427-608 or HESS J1708-410);

\item sources which show possible lower energies counterparts, but these lower energies counterparts are unidentified as well (such as HESS J1626-490);

\item sources which seemed to be of clear and quick identification, but deeper multiwavelength campaigns disproved those hypothesis (such as  HESS J1702-420, which seemed a clear example of middle-age PWN powered by the high spind-down luminosity pulsar PSR J1702-4128; however deeper X-ray campaigns disproved this idea);

\item unidentified sources which can be identified with deep multiwavelength campaigns (such as HESS J1731-347, which is the first SNR discovery triggered by TeV gamma-ray observations);

\item very extended sources which show loads of possible lower energies counterparts (such as HESS J1841-055 or HESS J1843-033, which might have very exotic counterparts as well);

\item very extended sources which, thanks to deeper IACTs observations, are discovered to be the convolution of several close-by sources instead (such as HESS J1745-303, which consists of three distinct sources);

\item and unidentified sources which might have very exotic possible counterparts.

\end{itemize}

\section{The discovery of HESS J1507-622 and Pulsar Wind Nebulae evolution}

The discovery of HESS J1507-622 obliged us to re-think about the time evolution of PWNe.
In fact HESS J1507-622 is a unique Galactic unidentified TeV gamma-ray source, very bright ($\sim$8\% of the Crab; 9.3 $\sigma$ of peak significance in9.7 hours of observations) and slightly extended, $\sim 0.16^{\circ}$ (H.E.S.S. collaboration 2011).
The uniqueness of HESS J1507-622 is given by its offset from the Galactic plane ($\sim 3.5^{\circ}$) and by the fact that it does not show any plausible counterparts at any wavelength. 
Given the brightness and the offset from the Galactic plane, it is really surprising to not find any counterpart especially at X-rays energies, because this Galactic latitude implies almost one order of magnitude less absorption than in the Galactic plane.

At IR and Radio wavelengths, there is no Southern Galactic Plane Survey and Spitzer GLIMPSE coverage of this part of the sky.
The Midcourse Space Experiment (MSX) and the MOLONGLO Galactic plane survey observed this region, but without evidencing any plausible counterparts.
It is located on a very long ($\sim 8^{\circ}$) radio filament at 2.4 GHz (Duncan et al. 1995). 
Looking into the complete CO survey (Dame et al. 2001), this H.E.S.S. source lies near the edge of a large ($\sim 5^{\circ} \times \sim 2^{\circ}$) nearby CO molecular cloud (moreover the
peak velocity of this cloud, around -5 km/s, would most likely place it quite near at a distance of $\sim 400$ pc); however, also in this case the substantial difference in extension and, in
the case of the CO molecular cloud, the offset of $\sim 1^{\circ}$ from the H.E.S.S. source centroid, suggested no obvious scenario for an association.

At X-rays it was very surprising to not find any X-ray counterpart, given the much lower absorption than in the Galactic plane. 
\emph{ROSAT} observations did not show any counterparts, but three near point sources were detected; in the case that one of them (especially 1RXS J150841.2-621006) could be discovered to be a pulsar, one can easily 
imagine an offset PWN scenario.
We obtained twice \emph{XMM-Newton} observations; the first observations (Proposal ID 05563102 - AO 7) were very severely affected by soft proton flare and so 35 ks re-observation has been obtained ( Proposal ID 06516201 - AO 9 ).
We obtained Chandra observations ( Proposal ID 10400599 - AO 10 ) as well and twice a deep (80 ks) Suzaku campaign: 80 ks the first time (priority C) and, even deeper, 120 ks the second, a couple
of years ago later. More details on the deep X-ray campaigns can be found in Tibolla et al. 2014.
These X-rays observations evidenced that the source 1RXS J150841.2-621006 / CXOU J150850.6-621018 slightly extended over the PSF of the instrument and it is coincident with the radio source MGPS J150850-621025;
this source could be either a SNR or a PWN (Tibolla et al. 2014), however cannot be a plausible counterpart of HESS J1507-622.

The absence of counterparts, especially in X-rays, would suggest a purely hadronic scenario. However hadronic scenarios would appear very disfavored, unless we would place HESS J1507-622 at a very small distance (less than 1 kpc!), since the density of target material off
the plane is very low; and, given the very small absorption at this Galactic latitude, it would be really challenging to explain the absence of lower energies counterparts.

The offset from the Galactic plane (could imply a low density and so) would suggest a leptonic scenario. 
A PWN powered by a very old pulsar could be a possible explanation for this source (in this case a small distance of the object is disfavored and HESS J1507-622 had to be placed at 6 kpc or farther).
This particular scenario was called Ancient PWN scenario (H.E.S.S. collaboration 2011; de Jager et al. 2009) and Ancient PWNe are indeed a recent idea that could explain a large fraction of the TeV gamma-ray unidentified sources.

In a scenario where the magnetic field decays as a function of time, the synchrotron emission will also fade as the PWN evolves.
And the B-field is indeed expected to decay as a function of time and the power-law index of the decay of the average nebular field strength has been calculated by means of MHD simulations (Ferreira \& de Jager, 2008).

Hence, while the synchrotron emission is fading as the PWN evolves, the very high energy (VHE,$E>10^{11}$ eV) emission depends on the CMB radiation field, which is constant on timescales relevant for PWN evolution.
For timescales shorter than the inverse-Compton lifetime of the electrons ($t_{IC} \simeq 1.2 \times 10^6 (E_e$/1 TeV)$^{-1}$ years), this will result in an accumulation of VHE
electrons which will also lead to an increased gamma-ray production due to up-scattering of CMB photons.
Such accumulation of very-high energy electrons in a PWN has indeed been observed in several VHE PWNe (such as the source HESS J1825-137).

This idea seems supported by the fact that the VHE PWNe sizes generally increase with pulsar age while the X-ray PWNe sizes show the opposite trend.
Moreover, for pulsars older than $\sim 10^3$ years the VHE PWNe are typically 100$-$1000 times larger than the sizes of the X-ray PWNe, while the
difference is only a factor 2 for some younger PWNe, like the Crab Nebula (e.g. Kargaltsev \& Pavlov 2010).
To summarize, during their evolution, PWN may appear as gamma-ray sources with only very faint low-energy counterparts and this may represent a
viable model for many unidentified TeV sources.
This effect can be clearly seen in the several examples shown by Okkie de Jager in 2008 and published in Tibolla et al. 2011b, where this model was applied to HESS J1507-622 to model for the first time the full spectral energy distribution of this unique TeV gamma-ray source.
The basic idea shown in those above-mentioned articles was developed into much more stringent models, such as a ``leaky box'' model (e.g. Tibolla et al. 2012) and our newest time
dependent PWN model (Vorster et al. 2013), obtaining comparable results.
In particular this latest model showed to be a particularily powerful tool to describe young PWNe systems, such as G21.5-0.9 (Vorster et al. 2013) and HESS J1420-607 (Kaufmann \& Tibolla 2017), as well as most of the unidentified TeV gamma-ray sources which have been treated by this model as aged/relic PWNe systems, such as:

\begin{itemize}

\item{HESS J1507-622 and HESS J1427-608 (Vorster et al. 2013);} 

\item{HESS J1837-069, HESS J1616-508, HESS J1702-420 and HESS J1708-410 (Tibolla et al. 2013);} 
 
\item{IGR J1849-0000 (Kaufmann \& Tibolla 2017).} 

\end{itemize}

Another important corollary of these long-living gamma-ray sources regards starburst galaxies: in fact, it has been
recently shown (Mannheim, Els\"asser \& Tibolla 2012) that PWNe are important and not negligible in explaining
the TeV gamma-ray emission detected from NGC 253 and from M82.
PWNe associated with core-collapse supernovae can readily explain the observed high TeV luminosities. The
final proof of this could arrive from deeper gamma-ray observations on other galaxies.

\section{Conclusions and final remarks}

Ancient PWNe are a very natural way to explain unidentified very high energy sources.
Moreover the ancient PWN models seem to be able to explain most of these unidentified sources (e.g. Vorster et. al. 2013, Tibolla et al. 2013, Kaufmann \& Tibolla 2017).
If indeed, at the termination shock of the Pulsar Wind also hadrons could be accelerated as well as leptons, as discussed in the first section, not only PWNe but also unidentified sources (i.e.
the dominant population at high energies and very high energies gamma-rays) can help in solving the CRs riddle.
We want to remark that, since in general PWNe Total Energy is comparable with SNRs Total Energy (or, at least, we can easily reach $\sim$10\% of it), if hadrons can really be
accelerated at the termination shock of Pulsar wind, this could indeed help to solve the riddle of the global picture of CR origin.






\references

Abeysekara, A. U., et al. : 2017, \journal{ApJ}, \vol{843}, id. 40.

Abdo, A. A. et al. : 2009, \journal{ApJ}, \vol{706}, L1.

Abdo, A. A. et al. : 2010, \journal{Science}, \vol{327}, 1103.

Abdo, A. A. et al. : 2010b, \journal{ApJ}, \vol{710}, L92.

Abdo, A. A. et al. : 2010c, \journal{ApJ}, \vol{718}, 348.

Abdo, A. A. et al. : 2010d, \journal{ApJ}, \vol{722}, 1303.

Abdo, A. A. et al. : 2011, \journal{ApJ}, \vol{734}, id. 28.

Ackermann, M., et al. : 2012, \journal{ApJ}, \vol{753}, id. 83.

Ackermann, M. et al. : 2013, \journal{Science}, \vol{339}, 807.

Aharonian, F. A., and Atoyan, A. M. : 1999, \journal{A\&A}, \vol{351}, 330. 

Aharonian, F. A., et al. : 2006, \journal{ApJ}, \vol{636}, 777.

Ajello, M. et al. : 2012, \journal{ApJ}, \vol{744}, id. 80.

Araudo, A. T., and del Valle, M. V. : 2014, \journal{Nucl. Phys. B}, \vol{256}, 117.

Atoyan, A. M., and Aharonian, F. A. : 1996, \journal{MNRAS}, \vol{278}, 525.

Baade, W., and Zwicky, F. : 1934, \journal{PNAS}, \vol{20}, 259.

Bednarek, W., and Protheroe, R. J. : 1997, \journal{PRL}, \vol{79}, 2616.

Bednarek, W., and Bartosik, M. : 2003, \journal{A\&A}, \vol{405}, 689.

Bednarek, W., Pabich, J., and Sobczak, T. : 2014, \journal{Nucl. Phys. B}, \vol{256}, 107.

Berezhko, E. G., and V\"olk, H. J. : 2006, \journal{A\&A}, \vol{451}, 981.

Cheng, K. S., et al. : 1990, \journal{J. Phys. G}, \vol{16}, 1115.

Dame, T. M., et al. : 2001, \journal{ApJ}, \vol{547}, 792.

de Jager, O. C., et al. : 2009, arXiv:0906.2644.

Drury, L. O'C.,  Markiewicz, W., and V\"olk, H. J. : 1989, \journal{A\&A}, \vol{225}, 179.
  
Drury, L. O'C.,  Aharonian, F. A., and V\"olk, H. J. : 1994, \journal{A\&A}, \vol{287}, 959.

Duncan, A. R., et al. : 1995, \journal{MNRAS}, \vol{277}, 36.

Ferreira, S. E. S., and de Jager, O. C. : 2008, \journal{A\&A}, \vol{478}, 17.

Ginzburg, V. L., and  Syrovatskii, S. L. : 1964, \journal{The Origin of Cosmic Rays}, authorised English translation by H. S. H. Massey, Pergamon Press, Oxford.

Giordano, F. et al. : 2012,  \journal{ApJ}, \vol{744}, id. L2. 

GLAST Science Brochure : 1996; http://glast.gsfc.nasa.gov.

Hartman, R. C., et al. : 1999, \journal{ApJSS}, \vol{123}, 79.

Hess, V. F. : 1912, \journal{Phys. Z.}, \vol{13}, 1084.

H.E.S.S. Collaboration : 2011, \journal{A\&A}, \vol{525}, id.A45.

Kargaltsev, O., and Pavlov, P. O. : 2010, \journal{AIP Conf. Ser.},  \vol{1248}, 25.

Katagiri, H. et al. : 2011, \journal{ApJ}, \vol{741}, id.44.

Kaufmann, S., and Tibolla, O. : 2017, accepted for publication in \journal{Nuclear and Particle Physics}.

Kolh\"orster, W. H. G. : 1914, \journal{Phys. Z.}, \vol{16}, 719.

Kotera, K. : 2014, \journal{Nucl. Phys. B}, \vol{256}, 131.

Mannheim, K., Els\"asser, D., and Tibolla, O. : 2012, \journal{Astroparticle Physics}, \vol{35}, 797.

Mannheim, K. : 2014, \journal{Nucl. Phys. B}, \vol{256}, 264.

Meszaros, P. : 2014, \journal{Nucl. Phys. B}, \vol{256}, 241.

Pacini, D. : 1912, \journal{Nuovo Cimento}, \vol{3}, 93.

Slane, P. et al. : 2014,  \journal{ApJ}, \vol{783}, id. 33.

Tam, P. H., et al. : 2010, \journal{A\&A}, \vol{518}, id.A8.

Tanaka, T. et al. : 2011,  \journal{ApJ}, \vol{740}, id. L51.

Thoudam, S. : 2014, \journal{Nucl. Phys. B}, \vol{256}, 125.

Tibolla, O. : 2007, \journal{MPLA}, \vol{22}, pp. 1611-1619.

Tibolla, O. et al. : 2011; \journal{25$^{th}$ Texas Symposium on Relativistic Astrophysics - TEXAS 2010 proceedings}, arXiv:1106.1023.

Tibolla, O., et al. (in memory of de Jager, O. C.) : 2011b, arXiv:1109.3144.

Tibolla, O., et al. : 2012, arXiv:1201.2295.

Tibolla, O., et al. : 2013, arXiv:1306.6833.

Tibolla, O., Drury, L. O'C. : 2014, \journal{Nucl. Phys. B}, \vol{256}, 1.

Tibolla, O., Kaufmann, S., and Kosack, K. : 2014, \journal{A\&A}, \vol{567}, id.A74.

Vink, J. : 2012; arXiv:1206.2363.

V\"olk, H. J. , and Biermann, P. L. : 1988, \journal{ApJ}, \vol{333}, L65.

Vorster, M. J., Tibolla, O., Kaufmann, S., and Ferreira, S. E. S. : 2013, \journal{ApJ}, \vol{773}, id.139.
 
Weinstein, A. : 2014, \journal{Nucl. Phys. B}, \vol{256}, 136.

Wulf, T. : 1909, \journal{Phys. Z.}, \vol{10}, 155.





\endreferences

\end{document}